\documentclass[twocolumn,letterpaper,showpacs,amsmath,amssymb]{revtex4}

\usepackage{graphicx}
\usepackage{dcolumn}
\usepackage{bm}

\begin{document}

\title{Numerical Study of the Stress Response of\\
Two-Dimensional Dense Granular Packings}

\author{N. Gland, P. Wang and H. A. Makse}
\affiliation { Levich Institute and Physics Department\\
City College of New York, New York, NY 10031}

\begin{abstract}
We investigate the Green function of two-dimensional dense random
packings of grains in order to discriminate between the different
theories of stress transmission in granular materials. Our
computer simulations allow for a detailed quantitative
investigation of the dynamics which is difficult to obtain
experimentally. We show that both hyperbolic and parabolic models
of stress transmission fail to predict the correct stress
distribution in the studied region of the parameters space. We
demonstrate that the compressional and shear components of the
stress compare very well with the predictions of isotropic
elasticity for a wide range of pressures and porosities and for
both frictional and frictionless packings. However, the states
used in this study do not include the critical isostatic point for
frictional particles, so that our results do not preclude the fact
that corrections to elasticity may appear at the critical point of
jamming, or for other sample preparation protocols, as discussed
in the main text. We show that the agreement holds in the bulk of
the packings as well as at the boundaries and we validate the
linear dependence of the stress profile width with depth.
\end{abstract}


\maketitle

\maketitle


%
%

\section{Introduction}

The theoretical understanding of the structural and mechanical
properties of static granular assemblies is still an open issue
\cite{mehta}. Indeed, there is no consensus on how the stress
distribution should be expressed and how granular materials
respond to applied perturbations. Experimental measurements of
stress distributions in silos and sandpiles
~\cite{ADD1,ADD2,ADD3,ADD4} seem to be incompatible with a simple
elastic theory and have led to the development of new mechanical
formulations for granular matter where new assumptions are
proposed \cite{B95,CLMNW96}.

%
%
The equations of equilibrium for the stress tensor, $\sigma_{ij}$,
are
\begin{equation}
\vec{\nabla} \, \bar{\bar{\sigma}} = \rho \vec{g}, \label{stress}
\end{equation}
where $\rho$ is the density of the medium and $\vec g$ the
gravitational acceleration vector. From this equation we see that
the stress is indeterminate. For instance in 2D, it provides two
independent equations for the three components of the stress
tensor (the stress tensor is symmetrical, $\sigma_{xy} =
\sigma_{yx}$). Thus a third constitutive equation (the missing
equation) characterizing the behavior of the material is needed in
order to provide closure to the problem. In the elastic
formulation closure is provided by  the introduction of the strain
field and new constitutive relations relating the strain with the
strain (Hooke's law). This leads to elliptic equations
governing the stress distribution \cite{landau}.%
%

Unlike elasticity--- where the closure is provided by Hooke's law---
new granular
models propose closure relations involving only stress tensor
components \cite{B95}. Hyperbolic equations of the same
mathematical structure as a wave equation are thus obtained. As a
consequence, a load applied at the surface propagates along two
characteristic directions.

%
%
Another theoretical framework of stress transmission in granular
matter (the q-model) was introduced by Coppersmith {\it et al.}
\cite{CLMNW96}. It considers the transmission of interparticle
forces by a random redistribution of forces among nearest
neighbors. At the continuum scale ~\cite{C98}, the stress
components satisfy a parabolic equation of diffusion, expected
from the stochastic uncorrelated nature of the transmission
process.

%
%
Thus, different theories predict different stress distributions
according to the nature of the equations governing the stress. The
importance of resolving which of these models apply under which
circumstances is of major theoretical and practical interest in
order to reveal the real mechanical properties of static granular
assemblies.

%
%
In order to test the validity of the different models, a simple
experiment was proposed by de Gennes ~\cite{PGD99}: the
calculation of the Green function by applying a perturbation force
on a grain in the granular medium and measuring the resulting
stress field perturbation. The elastic and the new models
predictions are then confronted. In the most simple case of an
homogeneous isotropic material, the elastic framework predicts
that the vertical stress profile resulting from the perturbation
is a single peak centered below the perturbation with a width $W$
proportional to the depth $y$ from the perturbation.  In the
hyperbolic models, the stress profile presents a distinctive
double peak and in the parabolic models the profile is a single
peak as in the elastic prediction but with a width proportional to
$W \sim \sqrt{y}$.

%
%

The Green function problem has divided the granular community into
two. For some researchers, the problem is well settled by now, in
particular after the experimental work of Reydellet and Clement
\cite{RC01} and Geng \emph{et al.} \cite{G01} in favor of the
validity of elasticity theory to model the response function.
Their experiments indicate a single peak in the stress
distribution in $\sigma_{yy}$ in qualitative agreement with the
elastic framework. Moreover, \cite{ABGRCBC} shows that the elastic
framework can be further extended to anisotropic system. However,
other experiments \cite{raj,double,HTW01,BRE02,SILV02,MOU04,GOL02}
show evidence for the possible validity of alternative approaches:
\cite{raj} discusses the lateral diffusion of vertical stress
increments due to the local application of a force in a brick
wall; \cite{HTW01} and \cite{GOL02} focus on relatively small
scale response experiments and \cite{GOL02} on ordered lattices;
The study on isostatic networks of \cite{HTW01} and \cite{MOU04}
show deviations from elasticity. These groups argue that on
approaching the isostatic point the elasticity theory would break
down and new approaches need to be considered to describe the
stress response. On the other hand some groups \cite{Roux} argue
that in the isostatic limit the hyperbolic equations do not
describe the stress field. Recent work \cite{Wyart} suggest that
there exist a characteristic length above which elasticity is
valid and below anomalous behaviour of isostaticity should be
observed. Also, it was shown that the degree of ordering
\cite{MUE02,SPA03,BON03,OTT03} and the texture properties of
packings \cite{atmana} affect strongly the stress transmission.
Even though, there is general consensus that the elastic theory
may describe qualitatively the system, many discrepancies still
remain \cite{serero}. A microscopic study is needed to perform a
{\it quantitative} assessment of the validity of the different
theoretical approaches.  Simulations offer an optimum scenario to
compare the different theories and allow to obtain microscopic
information \cite{atmana,SIL02} and the exact system dynamics
\cite{somfai}, difficult to observe in experiments.

%
%
Here we perform MD simulations to gain more microscopic insight
into the mechanical response of granular matter. Our results
indicate that the elastic theory can describe the simulations
better than the other models. Indeed, we find a stress response
whose vertical component shows one peak. Furthermore we show that
elasticity predicts very accurately all components of the stress
tensor at all depths in the packing, for all the ranges of
pressures, porosities and frictional properties studied in this
work.

%
\begin{figure}
\begin{minipage}[0]{1.\linewidth}
\centering { \resizebox{8cm}{!}{\includegraphics{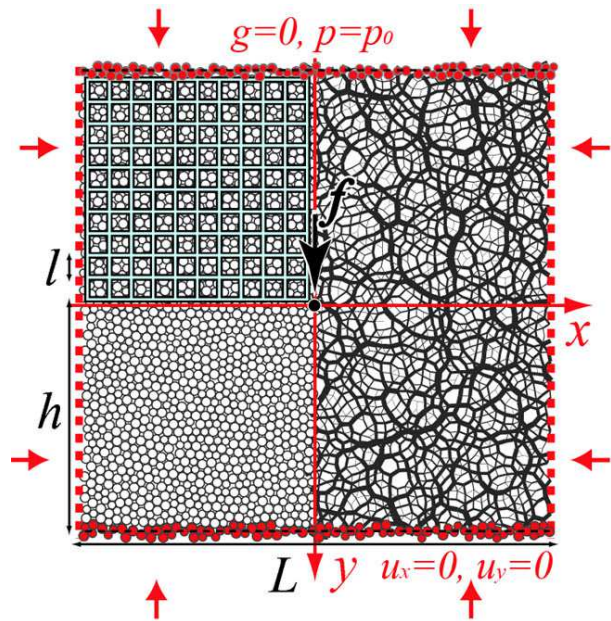}} }
%
\end{minipage}
\caption{System geometry. The left part of the figure shows a
detail of the bidisperse particle packing and the typical grid
used to calculate the coarse-grained stress profile, while the
right part shows the heterogeneity in the interparticle forces
(``force chains''). A small force $f$ is applied on a single grain
located at the center of the sample (origin of reference axes).
The vertical boundaries are composed of rigid grains while the
horizontal boundaries are periodic.}
\label{SYSTEM_GEOMETRY}
\end{figure}
%

\section{Numerical Simulations.}
%
%
 We developed a numerical code based
on the Discrete Element Method (DEM). The model deals
with an assembly of elastic spheres interacting via the
Hertz-Mindlin contact laws.
The grains interact with one another via non-linear Hertz normal
forces $f^{c}_{n}$ which depend on the overlap between two
spheres, and frictional transverse for\-ces $f^{c}_{t}$ which
depend both on the shear and normal displacements between the
grains:

\begin{equation}
\begin{array}{ll}
    f_{n}^{c}=\frac{2}{3} C_n R^{1/2}w^{3/2}, & \hbox{} \\
    \Delta f_{t}^{c}=C_t(Rw)^{1/2}\Delta s, & \hbox{} \\
\end{array}
\label{HERTZ-MINDLIN_CONTACT_LAWS}
\end{equation}
where $R=2R_{1}R_{2}/(R_1+R_2)$ with $R_1$ and $R_2$ the radii of the
spheres in contact. The normal overlap between the two grain is
$w$, and $\Delta s$ is the relative shear displacement between the
two grain centers. The constants $C_n=4 G /(1-\nu)$ and $C_t=4 G
/(2-\nu)$ are defined in terms of the shear modulus $G = 29$GPa
and the Poisson's ratio $\nu = 0.2$ of the material of the grains.
Coulomb friction, $f_t^c\le\mu_f f_n^c$,
\cite{johnson} is also included in the model and we set the
friction coefficient $\mu_{f} = 0.3$.  We also include viscous damping
terms in the equations of motions (translational and rotational)
to allow the system to relax toward static equilibrium.
In this study, the grains
spherical are constrained to move in 2D.

%
%
The packings are prepared using binary mixtures of particles at
equal concentrations characterized by a radius $R = (0.1 \pm
0.01)$mm in order to avoid crystallization.
%
%
We study different sample sizes ranging from $N=10000$ to
$N=250000$ particles
%
%
equilibrated using periodic boundary conditions in the horizontal
directions and two rigid walls made of grains at the top and the
bottom of the system (see Fig. \ref{SYSTEM_GEOMETRY}). In order to
prepare the larger samples ($N > 10000$) at static equilibrium and
save computational time on the preparation procedure, we repeated
periodically in space the elemental sample ($N=10000$)
equilibrated using periodic boundary condition in both horizontal
and vertical directions. For instance, to prepare a packing with
size $N=40000$, we repeat the elemental packing twice in both
horizontal and vertical directions. We found no significant
difference for large system size, so that most of the results
presented here are for $10000$ particles, which were obtained
without applying the repeating technique.
%
We use a numerical compaction protocol designed to prepare
confined dense granular isotropic packings explained in detail in
~\cite{MGJS04}. We set the gravity $\vec g = \vec 0$ since the
pressures of confinement considered in this work are sufficiently
high.

\begin{figure}
\begin{minipage}[0]{1.\linewidth}
\centering { \resizebox{8cm}{!}{\includegraphics{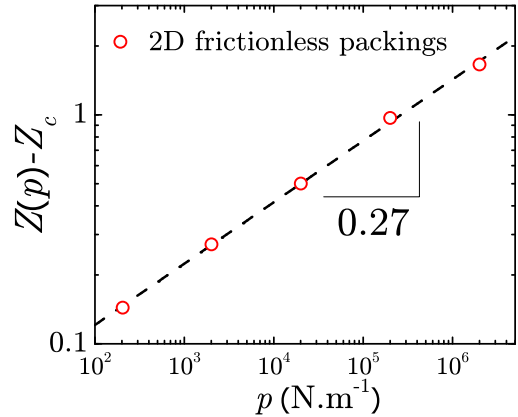}} }
\end{minipage}
\caption{Coordination number versus pressure obtained in 2D
frictionless packings, which follows the scaling behavior:
$Z(p)-Z_c \sim p^{0.27\pm0.07}$ with $Z_c=3.8$. Here the system is
isostatic with $Z \simeq 4$ as $p \rightarrow 0$. Note that the
critical coordination number $Z_c$ at isostatic limit of
frictionless case is actually a little smaller than 4, because of
the existence of floaters which carry no forces.}
\label{Coordination_Number}
\end{figure}

A packing is isostatic \cite{alexander} when the number of contact
forces equals the number of force balance equations. In a packing
of perfectly smooth (frictionless) particles, there are $NZ/2$
unknown normal forces and $DN$ force balance equations, where $D$
is the dimension and $Z$ is the average coordination number. This
result in a minimal coordination number needed for mechanical
stability as $Z_c=2D$, i.e., $Z_c=4$ and $6$ in 2D and 3D (See
Fig. \ref{Coordination_Number}, the system is at the isostatic
limit with $Z \simeq Z_c$ as $p \rightarrow 0$). In the case of
packings of perfectly rough particles, which is realized by
frictional particles with infinite friction $\mu\rightarrow\infty$
[note that in this case there are still tangential forces given by
the Mindlin elastic component Eq. 2], in addition to $NZ/2$
unknown normal forces and $DN$ force balance equations, there are
$(D-1)NZ/2$ unknown tangential forces and $D(D-1)N/2$ torque
balance equations \cite{edw}. Thus the coordination number in the
isostatic limit is $Z_c=D+1$, i.e., $Z_c=3$ and $4$ in 2D and 3D
for frictional packings. In such isostatic packings, there is
possibly a unique solution for the forces between particles for a
given geometrical configuration, because the number of equations
equals the number of unknowns. The existence of isostaticity is
the foundation of recent theories of stress propagation in
granular materials \cite{edw,TW,BB}.

Here we use a specific preparation protocol which sets the
friction between particles to zero. After a sample is equilibrated
at a given pressure, we consider two situations to calculate the
Green function: (i) frictionless case, $\mu = 0$; (ii) frictional
case, $\mu = 0.3$. Note that in (ii) friction is turned on an
equilibrated frictionless packing only for the calculation of the
Green function. Thus, this case is far from the isostatic
frictional case $Z_c=D+1$ \cite{Hepeng}. We use a frictionless
preparation because (1) we want to avoid issues of path dependence
in the sample preparations and (2) to obtain the isotropic
packings with large coordination number (6 in 3D, see
\cite{Bernal}, \cite{Chaikin}). Our preparation protocol tries to
mimic the shaking that is usually applied to the packings in
experiments to remove unstable voids that would otherwise render
the system unjammed. In a sense, the packings that we obtain are
analogous to the `` reversible" packings obtained in the
compaction experiments of Nowak {\it et al} \cite{Nowak}. This
preparation scheme was tested against experiments in our previous
computations of sound speeds in granular materials \cite{MGJS04}.
We find very good agreement between experiments and simulations.
Thus we believe that it is a reliable method to prepare the
packing. However it will be very interesting to see the response
behavior for a packing prepared with friction. For this case,
extreme care has to be taken to assure that the packing is jammed
and reversable, i.e., that there are no unstable voids. We have
recently developed a preparation protocol to provide jammed
frictional packings \cite{Hepeng}. The Green function calculation
will be applied to this frictional packings to investigate the
validity of elasticity as the frictional isostatic point is
reached.

%
%
We check the isotropy and randomness of the packings by
calculating both the texture tensor and the 2D
orientational order parameter ~\cite{RJMR96}.
%
%
In a first part, the simulated granular aggregates are prepared
with target pressure of $2.10^{3}$ N/m (equivalent to a pressure
of 10MPa in 3D) and result with an average coordination number $Z
\approx 4.3$ (see Fig. \ref{Coordination_Number}) and a solid
volume fraction $\phi \approx 0.860$. In a next part, in order to
be closer to the isostatic point, we show results for lower
pressures $2.10^{1}$ N/m and $2.10^{2}$ N/m (equivalent to 100KPa
and 1MPa in 3D) with $Z \approx 3.9$ and $4.1$ (see Fig.
\ref{Coordination_Number}) compared to the isostatic limit
$Z_c=4$, and $\phi \approx 0.838$ and $0.842$ slightly above the
random close packing limit of $\phi \approx 0.82$. The next step
of the preparation protocol consists to break the vertical
symmetry of the system by removing the periodic conditions in the
vertical direction and replacing it by rigid walls made of rough
particles at this time.

%
%

Next, we select the particle nearest to the center of the system
and apply a vertical downward force. The amplitude of the force,
$f = 10^{-6} \langle f \rangle$ (where $\langle f \rangle$ is the
average normal force in the packing), is small enough to assure
that we measure the linear response regime.
%
%
To calculate the local stress tensor $\sigma_{ij}(x,y)$, the
system is subdivided into square cells of size $l = 0.5$ mm
containing about $4$ to $5$ particles depending on their local
arrangements (see upper left corner of
Fig.~\ref{SYSTEM_GEOMETRY}). The results are independent of the
size of the coarse-graining cell in the range $l$ to $3 l$.
%
%

We have performed these simulations using three different system
sizes in a range $L$ to $3L$ in order to find the minimal size to
consider to avoid finite size effects. The aspect ratio $1$ to $3$
is needed for the appropriate study of the stress response
function because the width is of the order of $y$. We find that
the vertical response is less sensitive than the two others
component to the sample size. We have also compared
the vertical response functions above and below the perturbation,
respectively in the dilatation region and in the compression
region. We find an horizontal symmetry $\sigma_{ii}(x,y)
\approx \sigma_{ii}(x,-y)$, showing that the compression
response and the dilatation response are almost identical in the
studied linear regime.

We measure $(x,y)$ in units of $l$ with $x=0, y=0$ at the center
of the packing. We consider system sizes ranging from $L=40 l$
$(N=10000)$ to $L= 120 l$ $(N=90000)$.
We find no appreciable size effects
in this range for $\sigma_{yy}(x,y)$ and slight size effects for
$\sigma_{xx}(x,y)$ and $\sigma_{xy}(x,y)$ saturating above $L= 80
l$ $(N=40000)$.
%
%
The perturbation of the stress is calculated as $\sigma_{ij} =
\sigma_{ij}^{per}-\sigma_{ij}^{ref}$, where $\sigma_{ij}^{ref}$ is
the stress of the initial reference state before the perturbation
and $\sigma_{ij}^{per}$ is the stress after the perturbation is
applied ($\sigma_{ii} > 0$ corresponds to compression).
%
%
Because of the discrete nature of the material and the strong
inhomogeneity of the contact network \cite{RJMR96} (see
Fig.~\ref{SYSTEM_GEOMETRY}), we average the stress response
functions over many configurations of particles (about $20$).
%

%
\begin{figure}
\begin{minipage}[0]{1.\linewidth}
\begin{center}
%
\centering {
\resizebox{8cm}{!}{\includegraphics{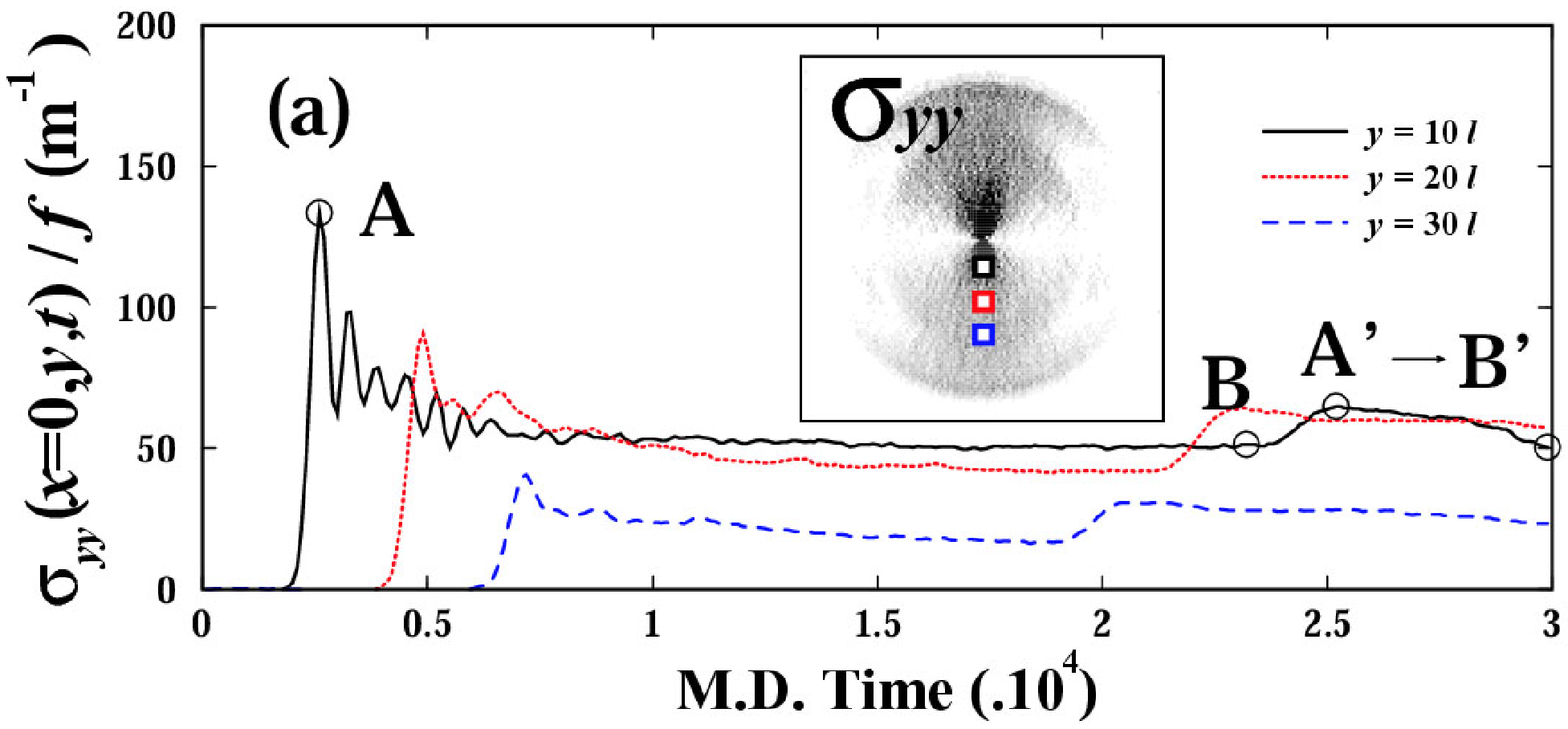}} }
%
%
%
%
%
\centering {
\resizebox{8cm}{!}{\includegraphics{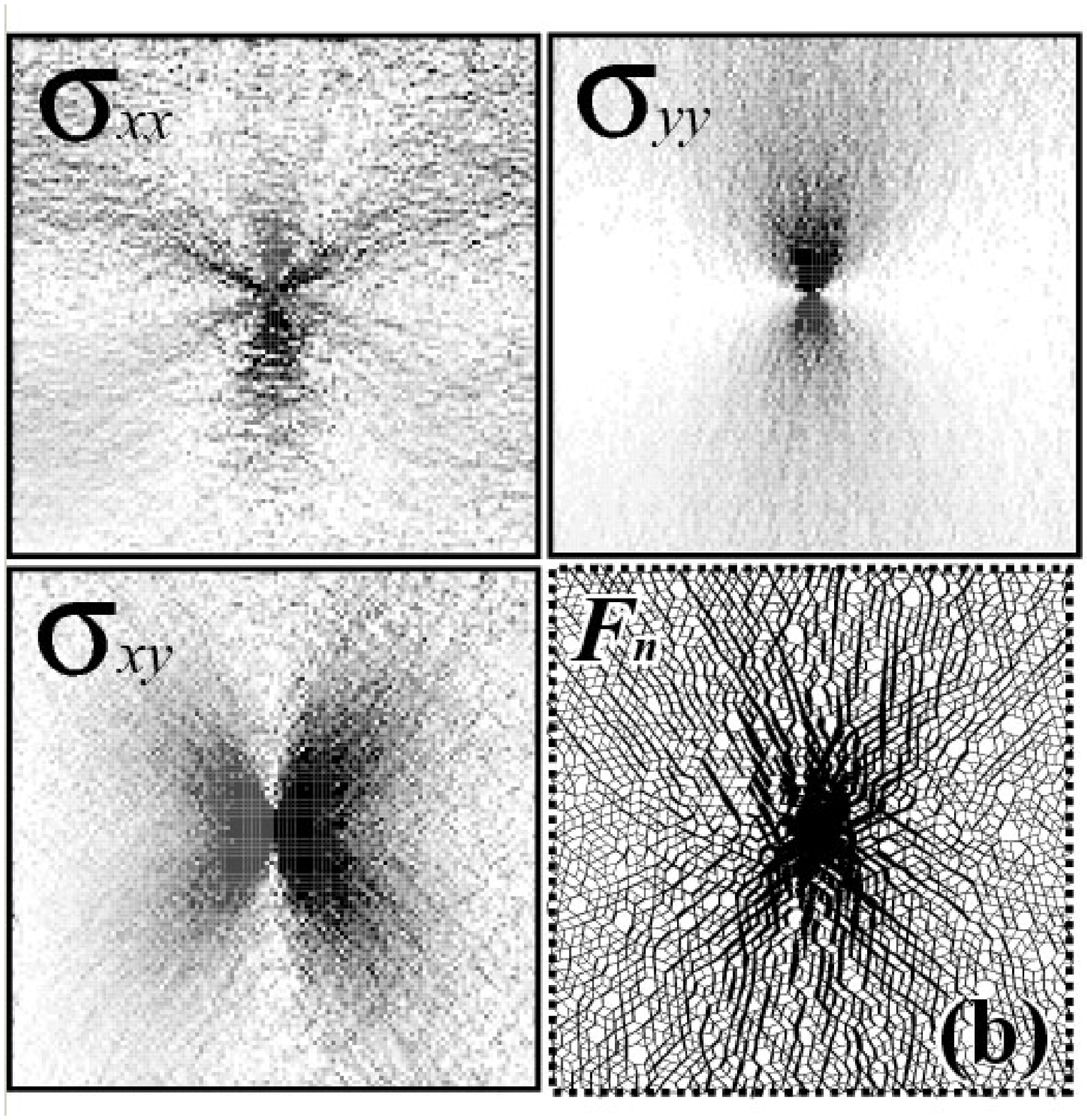}} }
\end{center}
\end{minipage}
\caption{(a) Time evolution of $\sigma_{yy}$ in a single
frictional packing at $x=0$ and at four different depths $y$ (as
indicated by the squares in the inset, showing the stress
propagation).
We observe an instantaneous stress peak (point A) in
the propagative regime as a result of the wave front first passage
followed by a relaxation to point B to the static regime and a
subsequent peak followed by another relaxation (from A' to B').
The slow damping drives the system from a propagative regime to a
static state.
(b) Typical stress response functions of one packing for the
three components of the stress tensor $\sigma_{xx}$, $\sigma_{yy}$
and $\sigma_{xy}$ and the enlarged detail of the perturbation of
the normal forces $F_n$ on the contact network shown in Fig.
\protect\ref{SYSTEM_GEOMETRY}.}
\label{GREEN_FONCTION_SIGMAIJ_AND_NORMALF}
\end{figure}
%
%
\section{Wave Propagation and Stress Distribution.}

The application of the force on the central particle generates a
pressure wave which propagates in the granular packing (as seen in
the inset of Fig. \ref{GREEN_FONCTION_SIGMAIJ_AND_NORMALF}a) and
is reflected at the rigid and periodic boundaries. The wave is
dissipated slowly over time until the packing reaches a new static
state at mechanical equilibrium. Figure
~\ref{GREEN_FONCTION_SIGMAIJ_AND_NORMALF}a presents the evolution
with time of the vertical stress $\sigma_{yy}(x=1,y,t>0)$ at
different depths $y$ below the perturbation in the compression
region. At all depths, a first stress maximum is observed at the
passage of the front wave (local instantaneous response in the
propagative regime, point A), followed by a rapid oscillation and
then a slow relaxation (point B). Then, a new stress peak of
smaller amplitude is measured (point A') which corresponds to the
wave reflected at the boundary followed by the same decrease
sequence towards complete relaxation to the new static state
(point B').
The further the point of measurement is, the smaller is
the amplitude of the first stress maximum resulting from the
expansion of the front.

 Figure ~\ref{GREEN_FONCTION_SIGMAIJ_AND_NORMALF}b
plots the stresses at equilibrium as well as the perturbation of
the normal interparticle forces, $F_n$.
We clearly see the existence of a single peak in
$\sigma_{yy}$ in agreement with elasticity and the experiments of
\cite{RC01,G01} and in disagreement with the predictions of the
hyperbolic theories of stress transmission.

%
\begin{figure}
\begin{minipage}[0]{1.\linewidth}
\begin{center}
%
\centering {
\resizebox{8cm}{!}{\includegraphics{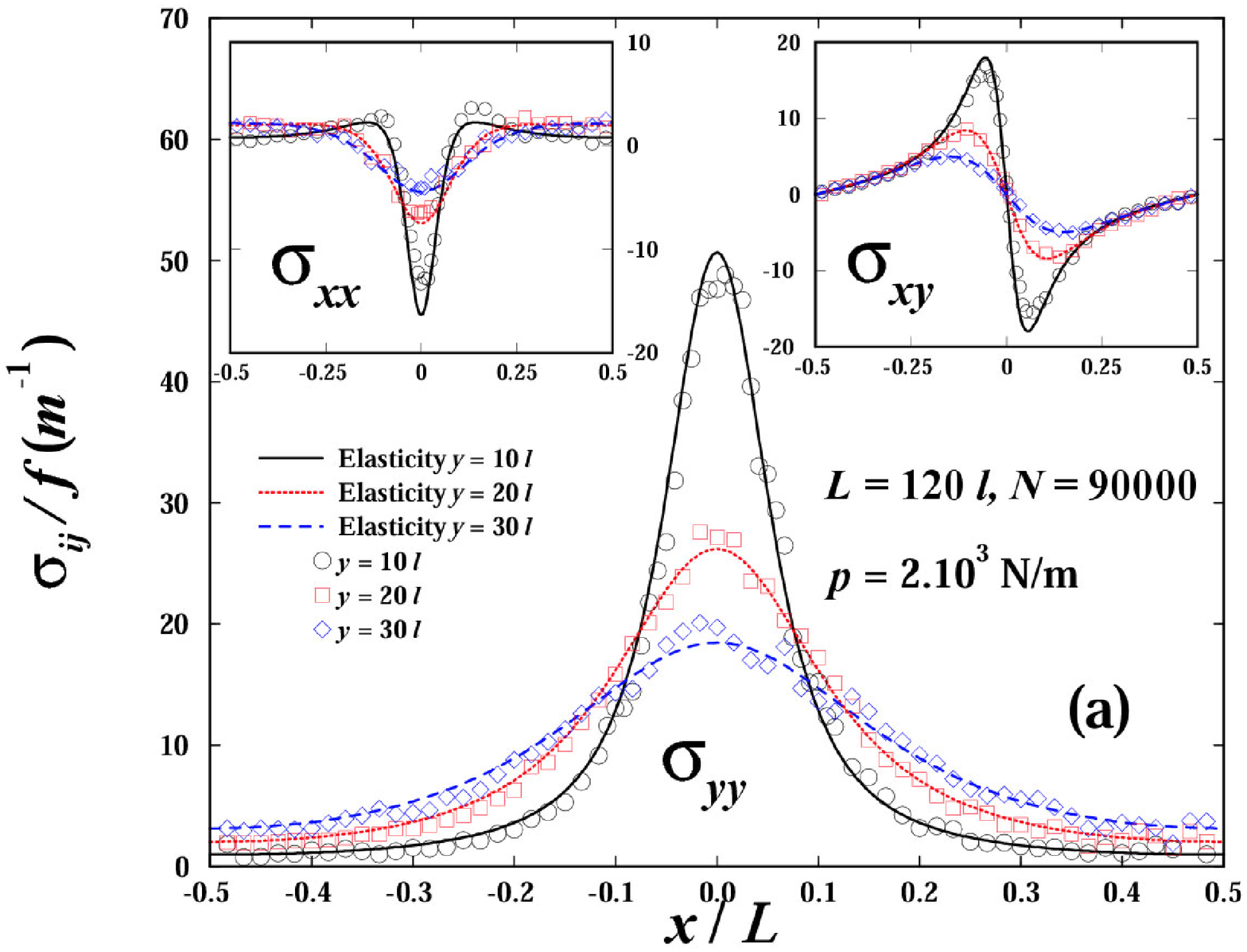}} }
%
%
%
%
\centering {
\resizebox{8cm}{!}{\includegraphics{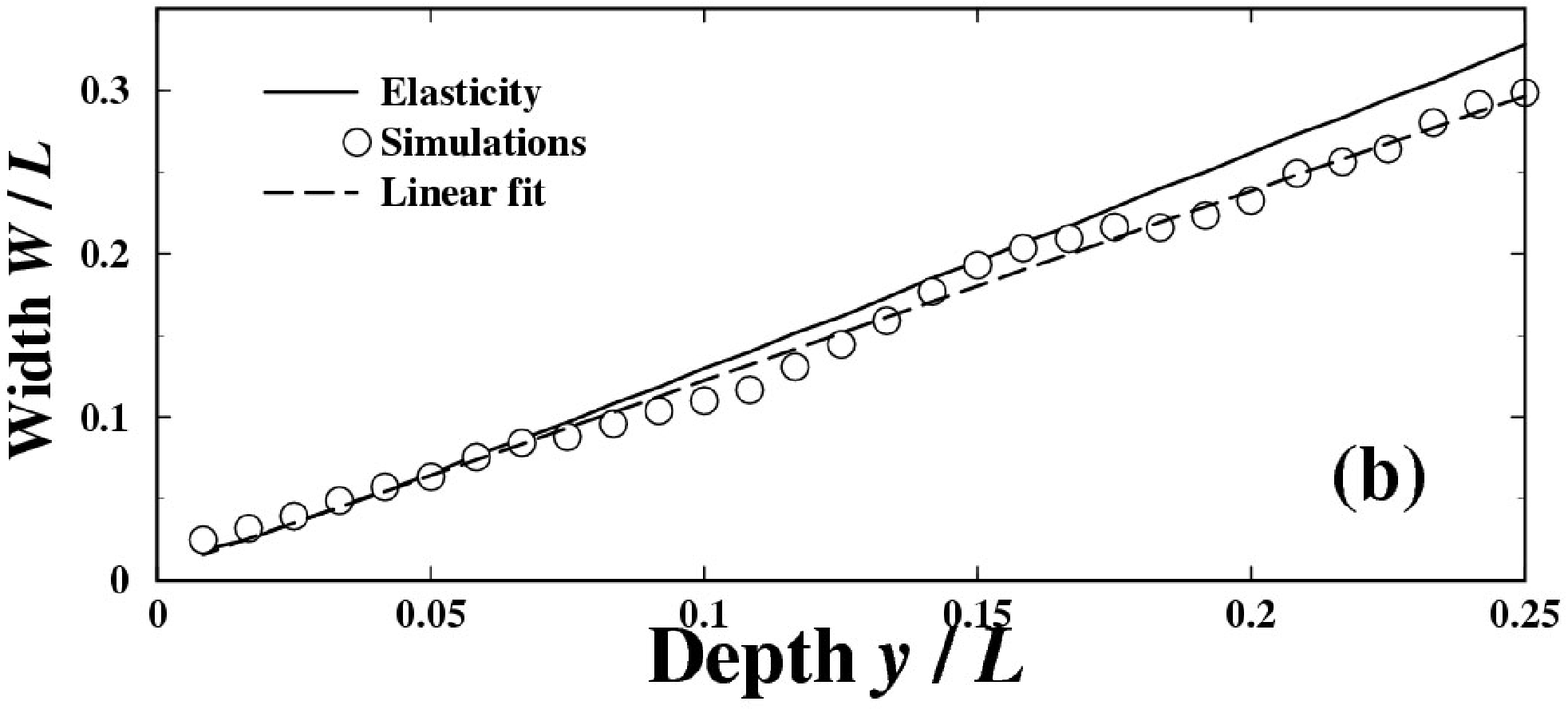}} }
%
\end{center}
\end{minipage}
\caption{(a) Comparison between the
solutions for the stress between simulations (symbols) and elasticity (lines)
 at different depths $y$ and at a fixed pressure. The
main plot compares the vertical component of the stress tensor
$\sigma_{yy}$ while the insets compare the two others components
$\sigma_{xx}$ and $\sigma_{xy}$.
(b) Half-width amplitude of the averaged vertical stress profiles
with increasing depth.
A linear dependence  is $W\sim 1.15 y$ is found, corroborating
the elastic solution $W\sim 1.30 y$.}
\label{STRESS_RESPONSE_FONCTIONS_ELASTICITY_SIMULATIONS}
\end{figure}
%
%
%

\section{Comparison with Elasticity.}

Next, we make a detailed comparison with the prediction of
elasticity theory. In the case of an infinite elastic medium the
solution was calculated by Boussinesq \cite{johnson}:
\begin{equation}
\sigma_{yy}(x,y) = \frac{2}{\pi} ~ \frac{f}{y}  ~\frac{1}
{\left(1+ (\frac{x}{y})^2\right)^{2}}.
\label{bousinesq}
\end{equation}
However, there is no analytical solution for the boundary
conditions used in the present part of the study, i.e., rigid top
and bottom rough boundaries where the strain satisfies  $u_{y}(x,y
= \pm L/2 ) = 0,$ and $u_{x}(x,y = \pm L/2) =0$ (this conditions
can be expressed in terms of the stress tensor as $\partial_{y}
\sigma_{xx}(x,y = \pm L/2) = (2+\nu) \partial_{x} \sigma_{xy}(x,y
= \pm L/2)$ and $\sigma_{xx}(x,y = \pm L/2) = \nu \sigma_{yy}(x,y
= \pm L/2)$ respectively, where $\nu$ is the Poisson ratio of the
entire packing \cite{serero}) and horizontal periodic boundary
conditions. Then, we find the solution of Eq. (\ref{stress})
numerically \cite{LTWB04}. The applied perturbation $Q(x)$ is
taken to be a narrow Gaussian centered at $x=0, y=0$, such that
$\sigma_{yy}(x,y=0,t>0)=Q(x)$.

In the theory, the stress distribution is not sensitive to the
Poisson ratio as we find numerically. Moreover, this result was
confirmed in previous simulations \cite{C98}. Since the results
are independent of $\nu$, a possible choice is to set the Poisson
ratio equal to the effective Poisson ratio, $\nu =\nu_e \simeq
0.28$, which is given by simulations \cite{MGJS04}. Moreover, the
effective Poisson ratio of a granular packing is independent on
pressure. [Note that some groups found the Poisson coefficients
varying with confining pressure, for frictionless packings at the
approach of the isostatic limit \cite{Hern} and frictional
packings with lower coordination number \cite{atmana}]. This can
be seen from the Effective Medium Theory calculations
\cite{MGJS04}, which gives:
$\nu_{e}=(K_e-2/3\mu_e)/(2K_e-2/3\mu_e)$, where $K_e$ and $\mu_e$
are the effective bulk modulus and shear modulus, respectively.
Due to the same power law relations of $K_e$ and $\mu_e$ versus
pressure, $K_e \sim p^{1/3}$ and $\mu_e \sim p^{1/3}$, the
effective Poisson ratio $\nu_e$ is independent of the pressure.
Therefore we use the same $\nu_e$ for all our systems at different
pressures. We note, however, that the results of \cite{MGJS04} are
for 3D while our systems are 2D. Other choices of $\nu$ would give
the same results as presented here. Figure
\ref{STRESS_RESPONSE_FONCTIONS_ELASTICITY_SIMULATIONS}a shows how
the elastic solution compares with the numerical results for the
stress components. The elastic solution provides a very
satisfactory fit at all depth for all the components of the stress
tensor.

\emph{Width of the Stress Profile.---}
We obtain the width of the vertical stress profile as a function of
the depth $y$ by calculating the half-amplitude spread of the
distribution ~\cite{RC01}. Figure
~\ref{STRESS_RESPONSE_FONCTIONS_ELASTICITY_SIMULATIONS}b shows
 a linear dependence on the depth, $W\sim y$, in general
agreement with elasticity (and in disagreement with the parabolic
model \cite{CLMNW96}). The numerical response ($W=1.15 y$)
exhibits a slightly narrower response than the elastic solution
($W=1.30 y$); it was also found in experiments ~\cite{RC01} and
could be attributed to the disorder of the packing.

%
\begin{figure}
\begin{minipage}[0]{1.\linewidth}
\begin{center}
%
\centering {
\resizebox{8cm}{!}{\includegraphics{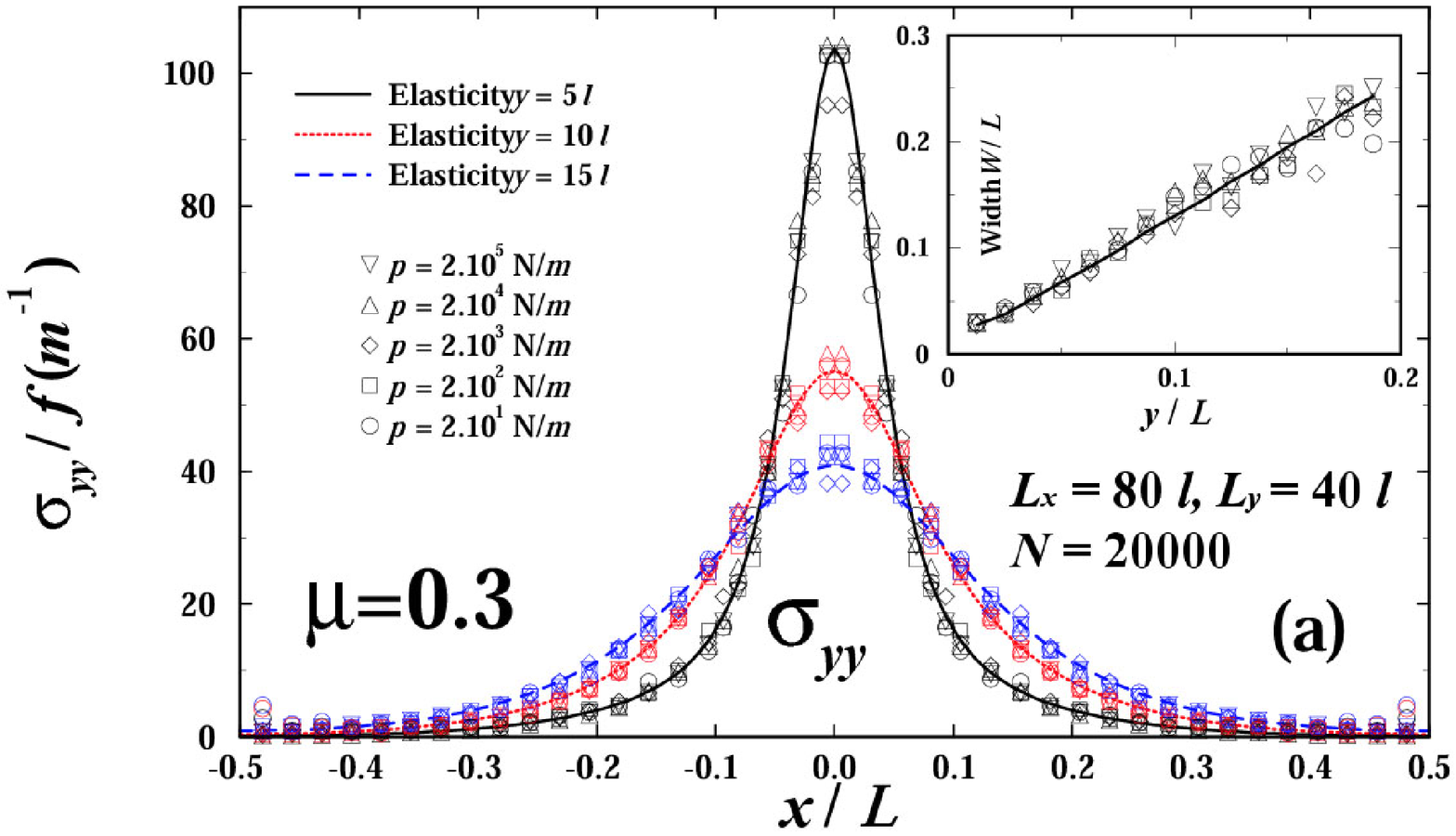}} }
%
%
%
%
%
\centering {
\resizebox{8cm}{!}{\includegraphics{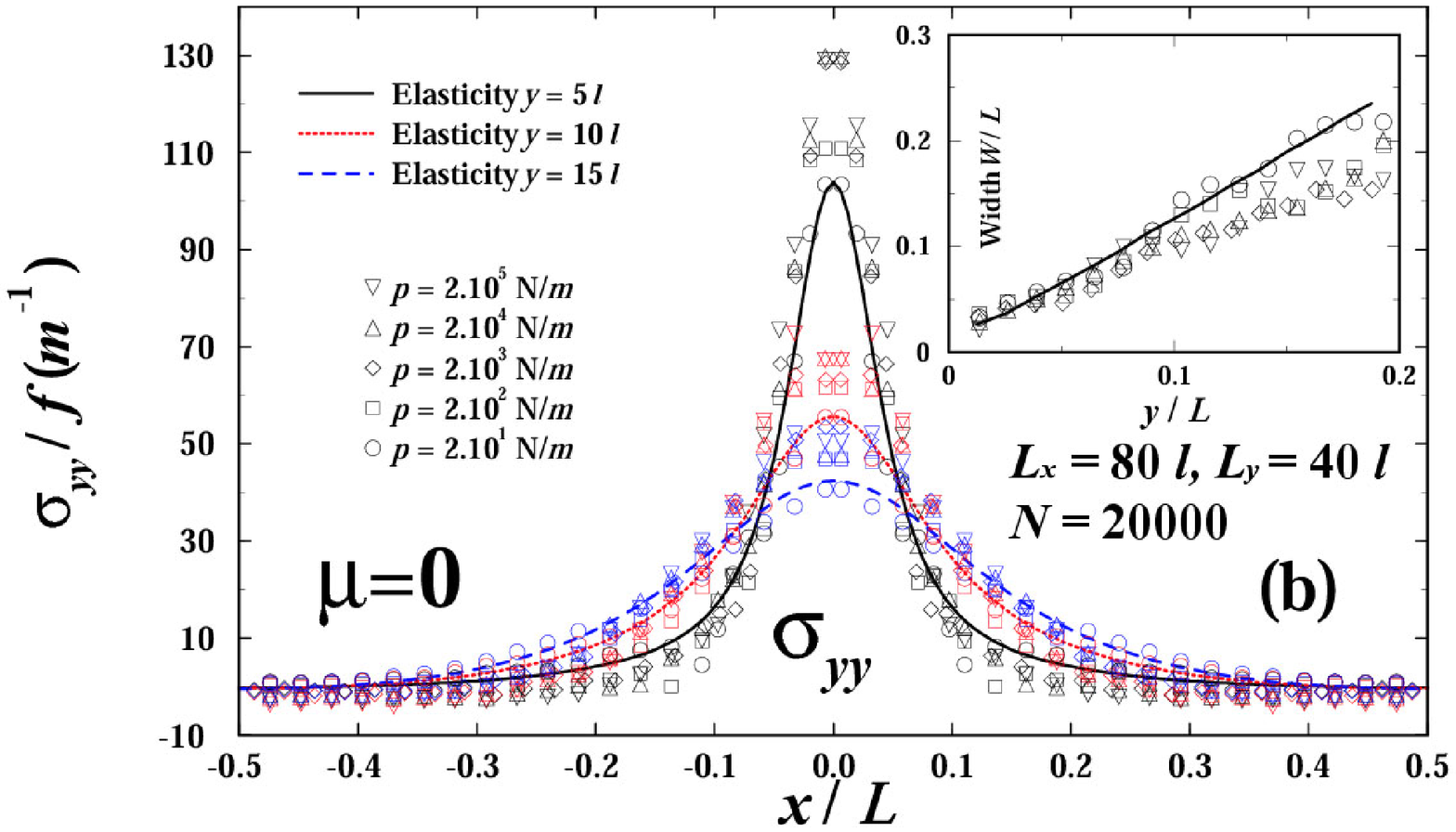}} }
%
\end{center}
\end{minipage}
\caption{Comparison between the vertical stress prediction of elasticity
and the simulations at different depths $y$ and for five
different pressures ranging from $p= 2.10^{1} N/m$ to $p =
2.10^{5} N/m$ for (a) rough boundaries conditions / frictional particles, (b)
smooth boundaries conditions / frictionless particles.
The  insets show the
half-width amplitude of the corresponding stress profiles
with increasing depth.
Linear dependence $W\sim y$ is found.}
\label{GREEN_FUNCTION_PRESSURE_FRICTION_DEPENDENCE}
\end{figure}
%

\emph{Pressure and Friction Effects}.--- We also study the stress
response functions for different pressures (from $p=2.10^{1}$ N/m
to $p=2.10^{5}$ N/m) and porosities on smaller systems ($L = 80 l$
and $N=20000$) shown in Fig.
~\ref{GREEN_FUNCTION_PRESSURE_FRICTION_DEPENDENCE}a. We find that
the stress profiles averaged over $20$ realizations are in
extremely good agreement with the elastic prediction for rough
boundaries. Moreover the profiles do not show appreciable pressure
dependence. This is in further agreement with elasticity. Indeed
the elastic solution depends only on the Poisson ratio $\nu$ which
is constant with pressure ~\cite{MGJS04}. Moreover, we find the
same linear dependence with the depth ($W = 1.20 y$) for the
studied range of pressure (see inset of Fig.
~\ref{GREEN_FUNCTION_PRESSURE_FRICTION_DEPENDENCE}a). The
agreement with elasticity holds very well also for the two others
components.

We also investigate the effect of friction by performing
simulations with frictionless packings ($\mu_{f}=0$) in the same
range of pressure and with the same system size shown in Fig.
~\ref{GREEN_FUNCTION_PRESSURE_FRICTION_DEPENDENCE}b. This time, we
compare the stress profiles averaged over $50$ realizations (due
to much larger fluctuations) with the elastic prediction for
smooth boundaries (i.e., $\sigma_{xy}(x,y = \pm L/2)=0$). We find
that the stress profiles show a small pressure dependence and that
the agreement with the elastic prediction holds better for the
smallest pressure. The reason for the discrepancy could come from
the fact that even though the particles are frictionless, the
fixed grains at the top and bottom surface induce a roughness, the
effect of which is expected to be more important at high
pressures. Indeed despite non-frictional particles, a non-zero
shear stress is found at the boundary, whose amplitude increases
with pressure. The linear dependence of the half-amplitude show
similar dependence with pressure.


\section{Summary}

Our numerical work shows that, for the studied
frictional and frictionless packings, the vertical component of
the stress response has a single peak at all depths and that the
half-amplitude width of the vertical stress profiles is
proportional to the depth. By calculating the stress profiles
using the elasticity framework and taking into account the exact
geometry of the simulations, we compare the numerical results and
find very good agreement with elasticity for the three
components of the stress tensor in the frictional case. Whereas,
experiments probe only the vertical stress profile at the
boundary, our numerical simulations are able to show the agreement
with elasticity at all depth inside the packings and all the
components. Furthermore, we confirm that pressure is not a
critical parameter for the stress response function of frictional
packings.

However, it is important to note that we have not investigate the
frictional isostatic point. Near this critical point the system
may show deviations from elasticity. Thus, we may still expect
corrections to appear if the system is prepared closer to the
isostatic frictional jamming point $Z_c=3$ in 2D. It will be of
interest, then, to continue this work and investigate the critical
regime as well. We also demonstrate that the response of
frictionless packings is consistent with elasticity; it exhibits
pressure dependency and tends to agree with elasticity at low
pressure. We attribute this effect to the roughness of the
boundaries in our simulations. In further work, it would be
interesting to confront the more general anisotropic elastic
theory to the response of loose and anisotropic packings obtained
by other compaction protocols.

\vspace{.5cm}

\section*{Acknowledgments}
We thank P. Claudin, E. Clement, Y. Gueguen, D. L. Johnson, E.
Kolb, D. Pisarenko and C. Song for stimulating discussions. This
work has been supported by the Department of Energy and the
National Science Foundation.



\end{document}